\documentclass[a4paper,review]{cas-sc}

\usepackage[numbers]{natbib}

\usepackage{amsthm}
\usepackage[T1]{fontenc}
\usepackage{caption}
\usepackage{graphicx}
\usepackage{float} 
\usepackage{subcaption}
\usepackage{multirow}
\usepackage[ruled,linesnumbered]{algorithm2e}

\usepackage{algpseudocode}
\usepackage{bm}
\usepackage{amsfonts}
\usepackage{amsmath}
\usepackage{array} 
\usepackage{enumitem}
\newtheorem{lemma}{Lemma}

\def\tsc#1{\csdef{#1}{\textsc{\lowercase{#1}}\xspace}}
\tsc{WGM}
\tsc{QE}
\tsc{EP}
\tsc{PMS}
\tsc{BEC}
\tsc{DE}
\begin{document}
\let\WriteBookmarks\relax
\def\floatpagepagefraction{1}
\def\textpagefraction{.001}

\shorttitle{A General Causal Inference Framework for Cross-Sectional Observational Data}

\shortauthors{Yonghe Zhao et~al.}

\title [mode = title]{A General Causal Inference Framework for Cross-Sectional Observational Data}                      
% \tnotemark[1,2]

% \tnotetext[1]{This document is the results of the research
%    project funded by the National Science Foundation.}

% \tnotetext[2]{The second title footnote which is a longer text matter
%    to fill through the whole text width and overflow into
%    another line in the footnotes area of the first page.}

\author[1]{Yonghe Zhao}[orcid=0000-0000-0000-0000]

%\cormark[1]


\ead{yhzhao21@mails.jlu.edu.cn}
\ead[URL]{https://www.zhihu.com/people/chebyshev-27}
\ead[URL]{https://s.dl100.cc/SuO6Cb}
%\ead[url]{www.cvr.cc, cvr@sayahna.org}

\credit{Conceptualization, Formal analysis, Methodology, Writing - original draft}

\affiliation[1]{organization={School of Artificial Intelligence, Jilin University},
    addressline={2699 Qianjin Street}, 
    city={Changchun},
    % citysep={}, % Uncomment if no comma needed between city and postcode
    postcode={130012}, 
    state={Jilin},
    country={China}}

% \author[1]{Qiang Huang}[style=chinese]

% \ead{huangqiang18@mails.jlu.edu.cn}

% \credit{Methodology, Writing - review \& editing}

% \author[2]{Shuai Fu}[style=chinese]

% \ead{fushuai@buaa.edu.cn}

% \credit{Data curation, Validation}

% \affiliation[2]{organization={State Key Laboratory of Software Development Environment, Beihang University},
%     addressline={37 Xueyuan Road}, 
%     city={Haidian},
    % citysep={}, % Uncomment if no comma needed between city and postcode
%     postcode={10019}, 
%     state={Beijing},
%     country={China}}

\author[1]{Huiyan Sun}[style=chinese]
\cormark[1]
% \fnmark[1,3]
%\ead{huiyansun@jlu.edu.cn}
%\ead[URL]{https://www.researchgate.net/profile/Huiyan-Sun}

\credit{Writing - review \& editing, Supervision}

%\affiliation[2]{organization={Sayahna Foundation},
%    city={Jagathy},
    % citysep={}, % Uncomment if no comma needed between city and postcode
%    postcode={695014}, 
%    state={Trivandrum},
%    country={India}}

%\author%
%[1,3]
%{Rishi T.}
%\cormark[2]
%\fnmark[1,3]
%\ead{rishi@stmdocs.in}
%\ead[URL]{www.stmdocs.in}

%\affiliation[3]{organization={STM Document Engineering Pvt Ltd.},
%    addressline={Mepukada}, 
%    city={Malayinkil},
    % citysep={}, % Uncomment if no comma needed between city and postcode
%    postcode={695571}, 
%    state={Trivandrum},
%    country={India}}

\cortext[cor1]{Corresponding author}
%\cortext[cor2]{Principal corresponding author}

%\fntext[fn1]{This is the first author footnote. but is common to third
%  author as well.}
%\fntext[fn2]{Another author footnote, this is a very long footnote and
%  it should be a really long footnote. But this footnote is not yet
%  sufficiently long enough to make two lines of footnote text.}

%\nonumnote{This note has no numbers. In this work we demonstrate $a_b$
%  the formation Y\_1 of a new type of polariton on the interface
%  between a cuprous oxide slab and a polystyrene micro-sphere placed
%  on the slab.
%  }

\begin{abstract}
Causal inference methods for observational data are highly regarded due to their wide applicability. While there are already numerous methods available for de-confounding bias, these methods generally assume that covariates consist solely of confounders or make naive assumptions about the covariates. Such assumptions face challenges in both theory and practice, particularly when dealing with high-dimensional covariates. Relaxing these naive assumptions and identifying the confounding covariates that truly require correction can effectively enhance the practical significance of these methods. Therefore, this paper proposes a General Causal Inference (GCI) framework specifically designed for cross-sectional observational data, which precisely identifies the key confounding covariates and provides corresponding identification algorithm. Specifically, based on progressive derivations of the Markov property on Directed Acyclic Graph, we conclude that the key confounding covariates are equivalent to the common root ancestors of the treatment and the outcome variable. Building upon this conclusion, the GCI framework is composed of a novel Ancestor Set Identification (ASI) algorithm and de-confounding inference methods. Firstly, the ASI algorithm is theoretically supported by the conditional independence properties and causal asymmetry between variables, enabling the identification of key confounding covariates. Subsequently, the identified confounding covariates are used in the de-confounding inference methods to obtain unbiased causal effect estimation, which can support informed decision-making. Extensive experiments on synthetic datasets demonstrate that the GCI framework can effectively identify the critical confounding covariates and significantly improve the precision, stability, and interpretability of causal inference in observational studies.
\end{abstract}

% \includegraphics{figs/grabs.pdf}

%\item Causal Reasoning and Diagnostics
%\item Observational Research
%\item De-confounding Artificial Intelligence

% Keywords

\maketitle

\section{Introduction}
In recent years, causal inference has attracted increasing attention across various domains, including epidemiology, healthcare, and economics\cite{NIPS2017_6a508a60,2016Causal,johansson2018learning}, etc. Compared to correlation, causality represents a more fundamental relationship between variables, revealing the directionality and determinacy\cite{imbens2015causal}. Randomized controlled trials (RCTs) are widely regarded as an effective means of exploring causality\cite{JudeaCausality}. However RCTs are time-consuming and expensive, even involving ethical issues in certain scenarios\cite{2010Reputation,2016Assessing,2011Unexpected}. How to conduct causal inference directly from collected observational data is a research topic of widespread concern.

Different from the randomness of treatments in RCTs, the main challenge of causal inference in observational studies is the unknown mechanism of treatment assignment\cite{imbens2015causal}. That is, there may be various deviations in observational data, for example the confounders that influence both treatment and outcome variables. Various causal inference methods for de-confounding are proposed based on the Potential Outcome Framework\cite{rubin1974estimating,splawa1990application}, such as reweighting\cite{rosenbaum1983central,2019Robust,lee2011weight,Austin2011An,imai2014covariate}, matching\cite{L2021Combining,2017Informative,JMLR21}, causal trees\cite{2010BART,Hill2011Bayesian,2017Estimation}, confounding balanced representation learning\cite{2018LearningWeighted,johansson2018learning,schwab2019perfect,2019AdversarialDu}, etc. However, these methods are commonly assumed that there exists the underlying hypothesis named as Confounding Covariates\cite{kuang2017treatment}. This hypothesis posits that all covariates act as confounders. The implementation of causal inference faces significant challenges in real-world scenarios due to the inability to adequately guarantee the assumption of Confounding Covariates, making it a topic of considerable attention and concern. Jessica et al.\cite{myers2011effects} validated that conditioning on instrumental variables introduces significant estimation bias through numerical experiments. Additionally, they emphasize the importance of distinguishing between confounding and instrumental variables. Several methods have relaxed the assumption of Confounding Covariates and incorporated pre-treatment variables in covariates. The pre-treatment variables include instrumental, confounding, and adjustment variables that are not affected by the treatment variables\cite{Rubin08}. The ${\rm D^{2}VD}$ model\cite{kuang2017treatment} was proposed as a data-driven approach to identify confounding and adjustment variables. In line with this, Tyler et al.\cite{vanderweele2019principles} advocated first identifying the causes of the treatment and outcome variables among the covariates before conducting causal inference\cite{vanderweele2019principles}. Essentially, these causes are equivalent to the pre-treatment variables.
Furthermore, Negar et al. devised a DR-CFR model\cite{hassanpour2019learning} that utilizes representation learning to obtain distinct representations of confounding, instrumental, and adjustment variables. Building upon this work, Kun Kuang et al. proposed the DeR-CFR model\cite{wu2020learning}, which extends the DR-CFR model by imposing additional constraints on instrumental variables to enhance the differentiation of these pre-treatment variables.
However, these methods have only studied one or a few components of the covariates and ensuring the fulfillment of the assumption of only pre-treatment variables in covariates remains elusive in practice\cite{ding2015adjust,vanderweele2011new}, particularly in scenarios involving high-dimensional covariates. Specifically, there is limited literature on the estimation bias caused by conditioning the post-treatment variables, such as mediators and colliders. When post-treatment variables are present, their interactions with pre-treatment variables may compromise the effectiveness of the aforementioned methods\cite{2000Causality}. 

Conducting causal inference for observational data continues to pose significant challenges: on one hand, the complexity of components within covariates make it difficult to generalize the assumption of Confounding Covariates for broader cases, thereby limiting its practical application; on the other hand, the intricate interplay among covariates makes it challenging to clearly define and identify all types of covariates, while the time-consuming nature of full causal diagram identification also presents a challenge. In response to these challenges, the aim of this paper is to establish a practical causal inference framework tailored for cross-sectional observational data, designed to eliminate confounding bias without introducing additional biases. Thus, a primary issue that must first be addressed is to concretely determine which key confounding covariates should be adjusted for and to strategically identify them. 

Addressing the aforementioned issues, this paper derives, from a theoretical standpoint based on the Markov properties of causal diagrams, that the key confounding covariates requiring adjustment are the common root ancestors of the treatment and the outcome variable. Subsequently, based on conditional independence characterization and causal asymmetry, a novel local graph identification algorithm ASI is developed for identifying these common root ancestors. Lastly, we integrate the ASI algorithm with de-confounding inference methods to construct a General Causal Inference (GCI) framework for cross-sectional observational data. The main contributions of this paper are as follows.
\begin{itemize}
\item We construct a GCI framework for cross-sectional observational data, which specifically targets key confounding covariates and mitigates their impact on causal inference. The GCI framework relaxes naive assumptions about covariates, rendering it of considerable practical significance.
\item This paper provides a theoretical analysis of key confounding covariates within complex covariates in observational studies. We also introduce a new local graph algorithm ASI to identify these key confounding covariates, which is grounded in solid theoretical foundations and has a satisfactory level of practical complexity.
\item Extensive experiments of the GCI framework alongside various state-of-the-art causal inference methods on synthetic datasets demonstrate that the GCI framework significantly enhances the precision and stability of causal inference. Moreover, compared to other methods, our framework offers superior interpretability.
\end{itemize}
\section{Related Works}
Due to the widespread practical significance of cross-sectional observational data research across various fields, a plethora of causal inference methods tailored for such data have emerged. These methods can be primarily categorized into two types: those based on the assumption of Confounding Covariates and those that relax this assumption to accommodate complex covariate compositions. In this sections, we provide a comprehensive overview of each approach.
\subsection{Causal Inference Methods based on the Assumption of Confounding Covariates}
\begin{figure}[ht]
\centering
\includegraphics[width=0.95\textwidth]{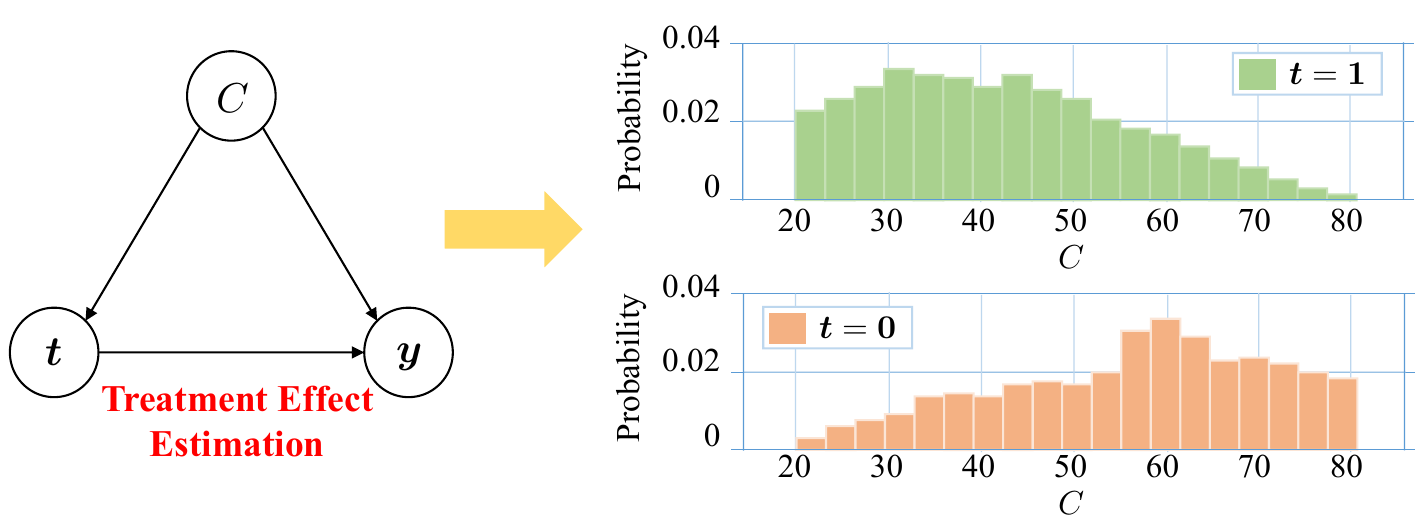}
\caption{The issues engendered by confounding covariates: inconsistent distribution of $C$ amidst discrete $t$ values.}
\label{confounders}
\vspace{-0.3cm}
\end{figure}

Methods based on the assumption of Confounding Covariates posit that all observed covariates are confounders, and thus their main objective is to address the challenge of confounding bias in causal inference. As shown in Fig. \ref{confounders}, the confounding covariates $C$ affect the selection of treatment $t$ thus leading to inconsistent distribution of $C$ among discrete $t$ values. Consequently, this phenomenon results in inaccurate counterfactual inference, which is similar to the domain adaptation problem\cite{yao2020survey}.

Therefore, the main objective of these methods is to achieve a balanced distribution of confounding covariates among different treatment groups\cite{yao2020survey}. These methods can be categorized into three categories based on the spatial domains: subspace of samples, sample space, and feature space. 
Methods based on the subspace of samples divide the samples based on certain balanced measures and then evaluate the causal effects in approximately balanced subspaces. Representative methods include Stratification\cite{2015Principal,hullsiek2002propensity}, Matching\cite{L2021Combining,2017Informative,JMLR21}, and the subspace partitioning methods based on decision trees, such as Bayesian Additive Regression Trees (BART)\cite{2010BART} and Random Causal Forest (RCF)\cite{2017Estimation}. Different from acquiring balanced subspaces, methods based on the sample space make weighted adjustments to the observational data of the original sample space to eliminate the influence of the confounders. 
The methods mainly include: (i) for discrete treatment variable: Inverse Propensity Weighting (IPW)\cite{rosenbaum1983central}, Doubly Robust estimator (DR)\cite{2019Robust}, Covariate Balancing Propensity Score (CBPS)\cite{imai2014covariate}; (ii) for continuous treatment variable: Inverse Conditional Probability-of-Treatment Weights (ICPW)\cite{2000Marginal,imbens2004nonparametric}, Boosting Algorithm for Estimating Generalized Propensity Scores (GBM)\cite{zhu2015boosting}, Covariate Balancing Generalized Propensity Score (CBGPS)\cite{2018Covariate}, etc. The weight-based adjustment methods depend on the rationality of constructing weights. 
Methods based on the feature space aim to acquire a balanced representation of the confounding covariates in an abstract representation space, which is crucial for performing downstream counterfactual inference tasks. 
Notable methods include the Treatment-Agnostic Representation Network (TARNet)\cite{shalit2017estimating}, Counterfactual Regression (CFR)\cite{shalit2017estimating}, and Local Similarity Preserved Individual Treatment Effect (SITE)\cite{yao2018representation} for discrete treatment variable, as well as DRNets\cite{2019LearningC} and the De-confounding Representation Learning (DRL)\cite{zhao2023deconfounding} model for continuous treatment variable.
\subsection{Methods for Identifying Complex Covariate Components}
The aforementioned causal inference methods assume the assumption of Confounding Covariates. However, practical considerations surrounding covariates often involve complex components that encompass both confounding and non-confounding variables\cite{zhao2023does}. This has sparked significant research interest in addressing these issues. 
Jessica et al.\cite{myers2011effects} presented the results of two simulation studies aimed at demonstrating that treatment effect estimate, conditioned on a perfect instrumental variable (IV) or a near-IV, may exhibit greater bias and variance compared to the unconditional estimate\cite{myers2011effects}. 
Kun Kuang et al. proposed a Data-Driven Variable Decomposition (${\rm D^{2}VD}$) algorithm\cite{kuang2017treatment} which automatically separate confounders and adjustment variables to estimate treatment effect more accurately. 
Tyler et al. put forward a practical approach\cite{vanderweele2019principles} for making confounder selection decisions based on the availability of knowledge regarding whether each covariate is a cause of the treatment or outcome variables. 
Negar et al. proposed the Disentangled Representations for Counterfactual Regression (DR-CFR) algorithm\cite{hassanpour2019learning} to identify disentangled representations of instrumental, confounding, and adjustment variables. Specifically, the DR-CFR algorithm constructs an objective function that relies on the independence between the representation of instrumental variables and the representation of confounding and adjustment variables. 
On this basis, Kun Kuang et al. constructed the Decomposed Representations for Counterfactual Regression (DeR-CFR) model\cite{wu2020learning}. This model introduces conditional independence constraints between the representation of instrumental variables and outcome variables to enhance the objective function of the DR-CFR algorithm. 
However, these methods make relatively naive assumptions about the covariates, as they have not comprehensively investigated the estimation bias caused by non-confounding covariates in causal inference, such as post-treatment variables.
\section{The General Causal Inference Framework for Cross-Sectional Observational Data}
\subsection{Problem Setting}
\subsubsection{Symbolic Description}
\begin{figure}[ht]
\centering
\includegraphics[width=0.55\textwidth]{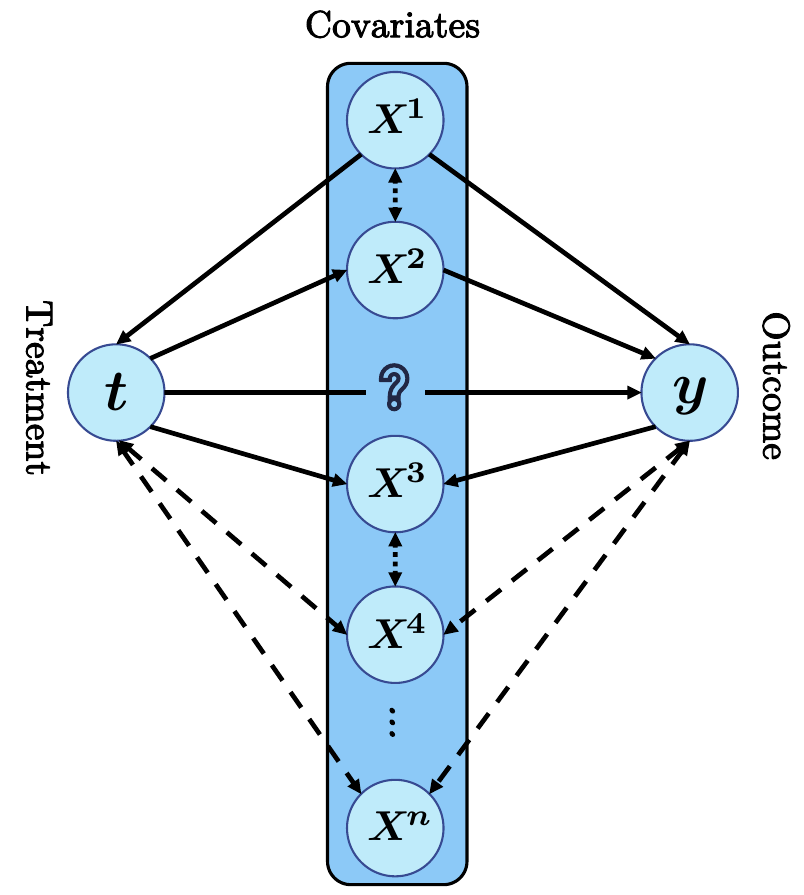}
\caption{The causal Diagram within the problem setting.}
\label{CASAUL}
\end{figure}

The GCI framework primarily aims at cross-sectional observational data, represented as $\mathcal{D} = \{X_{i},t_{i},y_{i}\}_{i = 1}^{N}$, which encompasses $N$ independent and identically distributed samples. The variable set $X=\{X^{i}\}_{i=1} ^{n}$ represents an n-dimensional covariates. It is worth noting that, in addition to confounding variables, $X$ may still contain various types of variables, such as instrumental variable, adjustment variable, mediating variable, collider variable, etc., as shown in Fig. \ref{CASAUL}. The specific nature of these covariates needs further determination in practical applications. The treatment variable $t \in \{0,1\}$ represents the control group ($t=0$) or the treatment group ($t=1$). Each value of the treatment variable, denoted as $t_{i}$, corresponds to a potential outcome indicated by $y(t_{i})$. These potential outcomes can be further categorized into the factual outcome $y^{f}$ and the counterfactual outcome $y^{\textit{cf}}$. Within the dataset $\mathcal{D}$, for the $i$-th sample, only the factual outcome $y^{f}(t_{i})$ corresponding to $t_{i}$ is accessible, while the other counterfactual outcomes $y^{\textit{cf}}(1-t_{i})$ remain unobserved. The primary objective of this paper is to infer counterfactual outcome $y^{\textit{cf}}$ by eliminating confounding bias without introducing additional biases. The emphasis lies in identifying the key confounding covariates that genuinely require adjustment for causal inference.
\subsubsection{Premise Hypothesis}
The GCI framework investigated in this paper require the fulfillment of three assumptions of the Potential Outcome Framework: stable unit treatment value (SUTV), unconfoundedness, and positivity assumption\cite{rosenbaum1983central,imbens2015causal}.

The SUTV assumption includes: firstly, the potential outcome of each individual is not affected by the treatment of any other individual, in other words, individuals are independent; secondly, there is no measurement error in the factual observational outcome. 

The unconfoundedness assumption represents that the treatment variable is independent of the outcome variable given the covariates $X$, i.e., $t\perp y|X$. With this unconfoundedness assumption, for the samples with the same covariates $X$, their treatment assignment can be viewed as random.

The positivity assumption, commonly referred to as the overlap assumption, posits that each value of $X$ can be assigned to any treatment with a non-zero probability, specifically $p(t|X=x) >0, \forall\ t, x$. The purpose of counterfactual inference is to assess differences across treatments, and the model is meaningless if some treatments can not be observed or are not meaningful. 
\subsection{Identifying Key Confounding Covariates based on the Markov Property of Causal Diagrams}
\subsubsection{The Markov Property of Causal Diagrams}
To identify the key confounding covariates, we begin by investigating the confounding variables that must be adjusted for causal inference based on the Markov properties, as shown in Lemma \ref{lemma1}, when a complete causal diagram is specified. This investigation will, in turn, inform the identification of the essential covariates.
\begin{lemma}
In a causal diagram, specifically referring to a Directed Acyclic Graph (DAG) in this paper, the distribution among the nodes satisfies the Markov property: For any given node, it is independent of all its non-descendant nodes, conditional upon its immediate parents.
\label{lemma1}
\end{lemma}
Based on the Lemma \ref{lemma1}, the joint distribution $P(t,y,X)$ of the nodes in the graph can be expressed as shown in Eq.~\eqref{joint_dis}.
\begin{equation}
\label{joint_dis}
P(t,y,X) = \prod_{\substack{i}} P(X^{i}|Pa(X^{i}))P(t|Pa(t))P(y|Pa(y))
\end{equation}

When an intervention $\textit{do(t)}$ is applied to the treatment variable, by setting $P(t = do(t)|Pa(t))=1$, the distribution after the intervention $P_{t}(y,X)$ can be expressed as Eq. \eqref{intervent_dis}, in accordance with the Markov property described in Eq. \eqref{joint_dis}.
\begin{equation}
\label{intervent_dis}
P_{t}(y,X)=\begin{cases}\prod_{\substack{i}} P(X^{i}|Pa(X^{i}))P(y|Pa(y)), & \text{intervention value } \textit{do(t)}\text{ is observable;} \\
0, &  \text{intervention value } \textit{do(t)}\text{ is unobservable;}\end{cases}
\end{equation}

Consequently, in the case where the intervention $\textit{do(t)}$ is observable, the desired treatment effect $P_{t}(y)$ can be obtained by integrating over the covariates, which is represented by Eq. \eqref{treatment_effect}. In the following section, the Eq. \eqref{treatment_effect} is utilized to guide the identification of key confounding covariates.
\begin{equation}
\label{treatment_effect}
P_{t}(y) = \sum_{\substack{X^{i}}}\prod_{\substack{i}} P(X^{i}|Pa(X^{i}))P(y|Pa(y))
\end{equation}

\subsubsection{The Derivation of Key Confounding Covariates}
It is evident from Eq. \eqref{treatment_effect} that calculating the causal effect $P_t(y)$ requires the adjustment of all covariates, which results in high complexity. To further clarify the key confounding covariates that truly need adjustment, we must perform an in-depth simplification and derivation of $P_t(y)$. 

\begin{figure}[ht]
\centering
\includegraphics[width=0.75\textwidth]{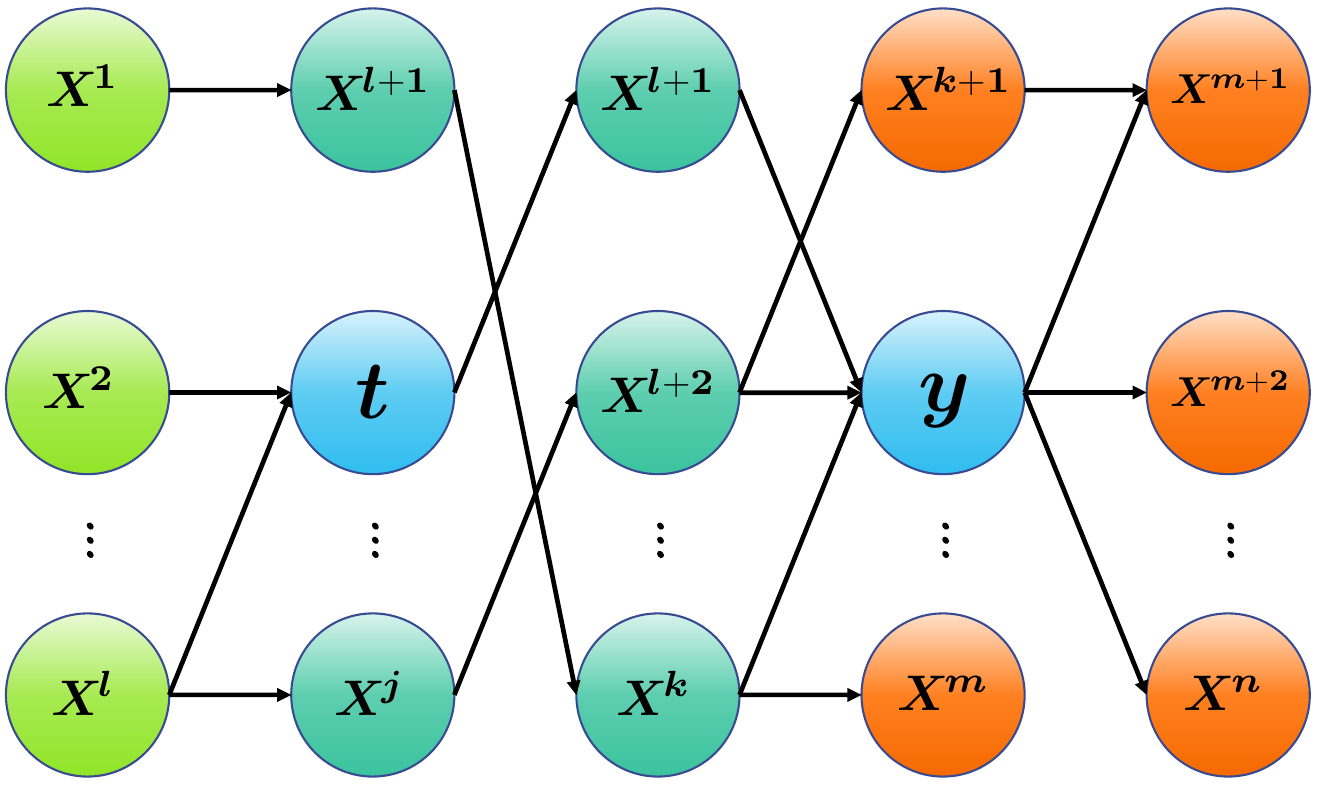}
\caption{The causal diagram unfolded according to the topological order (parent nodes before children nodes).}
\label{order}
\end{figure}

For ease of exposition, we unfold the causal diagram in Fig. \ref{CASAUL} according to the topological ordering of the diagram, and categorize the covariate nodes into three types: root ancestor nodes $X^{\textit{RA}}$ (green nodes), non-root ancestor nodes $X^{\textit{N-RA}}$ (cyan nodes), and non-ancestor nodes $X^{\textit{N-A}}$ (orange nodes), as shown in Fig. \ref{order}. The topological ordering of nodes on a causal diagram refers to an arrangement where parent nodes precede their child nodes in sequence.

Drawing on Fig. \ref{order}, the subsequent analysis delineates the roles of the three specified variable categories in the identification of the causal effect $P_{t}(y)$. Beforehand, Lemma \ref{lemma2} provides an exposition of the essential principles underpinning the simplification of $P_{t}(y)$.
\begin{lemma}
For the post-intervention distribution $P_{t}(y)$ shown in Eq. \eqref{treatment_effect}, if a variable $X^{i}$ is not a parent node of any other variable $X^{j}$ where $i\neq j$, then the variable $X^{i}$ can be integrated out from $P_{t}(y)$. In other words, during the computation of integrals over products of conditional probability terms, if a variable does not appear in the conditions of any conditional probability terms, then that variable can be integrated out. 
\label{lemma2}
\end{lemma}
Considering the variable $X^{n}$, which is positioned at the end of the topological sequence and does not serve as a parent node for any variable $X^{j}(j\neq n)$. The proof of Lemma 2 is elucidated in Equation \eqref{func_c_x}.
\begin{equation}
\label{func_c_x}
\begin{aligned}
P_{t}(y) &= \sum_{\substack{X^{i}}}\prod_{\substack{i}} P(X^{i}|Pa(X^{i}))P(y|Pa(y))\\
&= \sum_{\substack{X^{1}}}^{X^{n-1}}\prod_{\substack{i=1}}^{n-1} P(X^{i}|Pa(X^{i}))P(y|Pa(y))\underbrace{\sum_{\substack{X^{n}}}P(X^{n}|Pa(X^{n}))}_{=1} \\
&= \sum_{\substack{X^{1}}}^{X^{n-1}}\prod_{\substack{i}}^{n-1} P(X^{i}|Pa(X^{i}))P(y|Pa(y))\\
\end{aligned}
\end{equation}

Utilizing Lemma \ref{lemma2} and Fig. \ref{order} as a foundation, the derivation of $P_{t}(y)$ is systematically conducted through the subsequent steps:
%\begin{enumerate}
%  \item 
%\end{enumerate}
%\setlist[description]{leftmargin=*} % 设置所有description环境的缩进
\begin{description}
    \item[Step 1:] In accordance with Lemma \ref{lemma2}, by addressing non-ancestor nodes $X^{N-A}$ (orange nodes in Fig. \ref{order}) in a reverse sequence based on their topological ordering ($X^{n}\rightarrow X^{n-1}\rightarrow \cdots \rightarrow X^{k+1} $), it is readily observed that the nodes $X^{N-A}$ can be incrementally removed from Eq. \eqref{treatment_effect}, resulting in Eq. \eqref{treatment_effect_s1}.
    \begin{equation}
        \label{treatment_effect_s1}
        P_{t}(y) = \sum_{\substack{X^{A}}}\prod_{\substack{\{i:X^{i}\in X^{A}\}}} P(X^{i}|Pa(X^{i}))P(y|Pa(y))\ , X^{A} = \{X^{i}\}_{i=1}^{n}\setminus X^{\textit{N-A}}
    \end{equation}
    \item[Step 2:] Likewise, with respect to ancestor nodes $X^{\textit{tA}}$ that affect $y$ solely through $t$  (for instance, $X^2$ in Fig. \ref{order}), given that $P_{t}(y)$ lacks the conditional terms of treatment variables $P(t|Pa(t))$, it is possible to systematically exclude these nodes from $P_{t}(y)$ following a reverse topological sequence, thereby deriving Eq. \eqref{treatment_effect_s2}.
    \begin{equation}
        \label{treatment_effect_s2}
        P_{t}(y) = \sum_{\substack{X_{t}^{A}}}\prod_{\substack{\{i:X^{i}\in X_{t}^{A}\}}} P(X^{i}|Pa(X^{i}))P(y|Pa(y))\ , X_{t}^{A} = X^{A}\setminus X^{\textit{tA}}
    \end{equation}
    \item[Step 3:] For any non-root ancestor nodes $X^{\textit{N-RA}}$ (cyan nodes in Fig. \ref{order}), they can be effectively removed from the conditional terms within the conditional expressions by employing the transformation formula delineated in Eq. \eqref{transform}.
    \begin{equation}
        \label{transform}
        P \bigl(y|Pa\left(y\right) \bigr)P\left(Pa^{\textit{N-R}}(y)|Pa\left(Pa^{\textit{N-R}}(y)\right)\right) = P\left(y,Pa^{\textit{N-R}}(y)|Pa(y)\bigcup Pa\left(Pa^{\textit{N-R}}(y)\right)\setminus Pa^{\textit{N-R}}(y)\right)
    \end{equation}
    The Eq. \eqref{transform} demonstrates that for any non-root parent nodes $Pa^{\textit{N-R}}(y)$ related to the outcome variable $y$, a chain of conditional transformation can be applied, which results in the substitution of its parent node $Pa\left(Pa^{\textit{N-R}}(y)\right)$ into the position of its conditional terms.

    Building on Eq. \eqref{transform}, when all non-root ancestor nodes $X^{\textit{N-RA}}$ are transformed following a reverse topological sorting, they are effectively removed from the conditional terms of all conditional probabilities. At this point, one can apply Lemma \ref{lemma2} to exclude $X^{\textit{N-RA}}$ from $P_{t}(y)$, which facilitates the derivation of Eq. \eqref{treatment_effect_s3}.
    \begin{equation}
        \label{treatment_effect_s3}
        P_{t}(y) = \sum_{\substack{X_{t}^\textit{RA}}}\prod_{\substack{\{i:X^{i}\in X_{t}^\textit{RA}\}}} P(X^{i})P(y|X_{t}^\textit{RA})\ , X_{t}^\textit{RA} = X_{t}^{A} \setminus X^{\textit{N-RA}} = X^\textit{RA} \setminus X^{\textit{tA}}
    \end{equation}

    \item[Step 4:] For nodes $X_\textit{Nt}^\textit{RA}$ belonging to $X_{t}^\textit{RA}$ but not being ancestors of the treatment variable (for instance, $X^1$ in Fig. \ref{order}), these nodes can be removed from the conditional expression of $y$ by introducing a condition on $t$, as shown in Eq. \eqref{transform2}.
    \begin{equation}
    \label{transform2}
    \begin{aligned}
    P(y|X_{t}^\textit{RA})P(X_\textit{Nt}^\textit{RA}) &= P(y|X_{t}^\textit{RA})P(X_\textit{Nt}^\textit{RA}|t)\\
    &= P(y,X_\textit{Nt}^\textit{RA}|X_{t}^\textit{RA} \setminus X_\textit{Nt}^\textit{RA} \bigcup t )
    \end{aligned}
    \end{equation}
    Following this, leveraging Lemma \ref{lemma2} allows for the exclusion of any ancestor nodes $X_\textit{Nt}^\textit{RA}$ unrelated to treatment $t$ from $P_{t}(y)$, culminating in the formulation of Eq. \eqref{treatment_effect_s4}.
\begin{equation}
        \label{treatment_effect_s4}
        P_{t}(y) = \sum_{\substack{X^\textit{A*}}}\prod_{\substack{\{i:X^{i}\in X^\textit{A*}\}}} P(X^{i})P(y|X_{t}^\textit{RA})\ , X^\textit{A*} = X_{t}^\textit{RA} \setminus X_\textit{Nt}^\textit{RA}
    \end{equation}
\end{description}

To conclude, when the causal diagram is clearly defined, it is possible to ascertain, based on the Markov properties of the causal diagram (Lemma \ref{lemma1}) and Lemma \ref{lemma2}, that \textbf{the key confounding covariates requiring adjustment for the assessment of the causal effect $P_{t}(y)$ of $t$ on $y$ are limited to their common root ancestor nodes (excluding root nodes that affect $y$ only through $t$), denoted as $X^\textit{A*} = \textit{An}(t)\bigcap \textit{An}(y)\bigcap \textit{Root}(X)$.} Building on this conclusion, we have designed an Ancestor Set Identification (ASI) algorithm and developed a general causal inference framework for cross-sectional observational data.
\subsection{An Ancestor Set Identification Algorithm for Outcome Variables}
\begin{figure}
\centering
\includegraphics[width=0.98\textwidth]{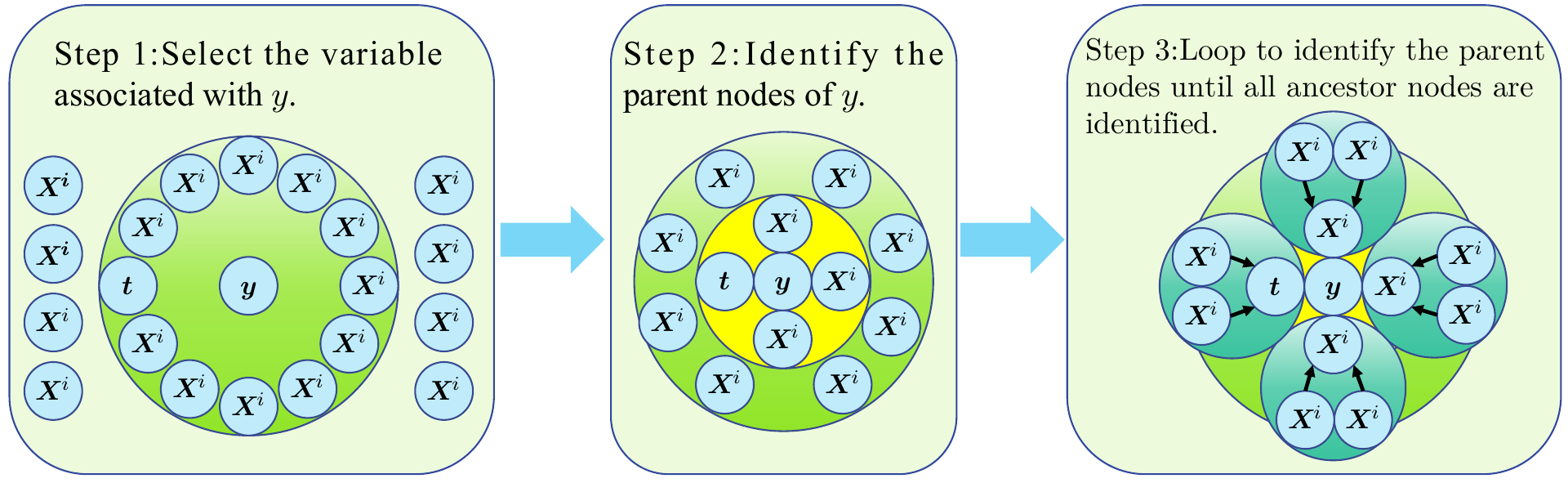}
\caption{The logical framework diagram of the ASI algorithm.}
\label{ans}
\end{figure}

Drawing from the conclusions reached earlier, the identification of the common ancestors $X^\textit{A*}$ shared by the treatment variable $t$ and the outcome variable $y$ is a prerequisite for engaging in causal inference. It becomes evident that, when $t$ is an ancestor of $y$, the set $X^\textit{A*}$ is encompassed within the ancestral set $\textit{An}(y)$ of $y$. Consequently, after ascertaining $\textit{An}(y)$, one can pinpoint $X^\textit{A*}$ based on its definition, thereby enabling the pursuit of an unbiased inference of causal effects. This section delineates an ASI algorithm designed to identify $\textit{An}(y)$, with the algorithm's rationale depicted in Fig. \ref{ans}.

As depicted in Fig. \ref{ans}, the proposed algorithm for identifying the ancestral set is an iterative process, whose primary function is to identify the parent nodes of any given variable, specifically divided into three steps:
\begin{enumerate}
  \item Filtering variables related to the target variable;
  \item Determining the parent and child nodes of the target variable based on conditional independence nature, and subsequently identifying the parent nodes in accordance with causal asymmetry;
  \item Iteratively conducting Steps 1 and 2 to identify the parent and ancestral nodes of the target variable, thereby achieving the identification of the target variable's ancestral set.
\end{enumerate}
This ASI algorithm constitutes a local graph identification algorithm, where the complexity of the algorithm is effectively reduced by recursively narrowing the scope of identification. The specific algorithmic procedures are presented in Algorithms \ref{AnsIdentify} and \ref{PCIdentify}.

Algorithm \ref{AnsIdentify} exemplifies the iterative step within the ASI algorithm. Specifically, lines 2-17 iteratively invoke the parent and child nodes identification algorithm shown in Algorithm \ref{PCIdentify} to ascertain the ancestral set of the target variable. Notably, in addition to logging parent nodes (line 6), Algorithm \ref{AnsIdentify} concurrently documents the child nodes of the target variable during each iteration  (line 7) and excludes these nodes in subsequent cycles (line 13) to reduce the algorithm's complexity. This approach is predicated on the acyclic assumption in the graph.

\begin{algorithm}
    \caption{The algorithm procedure of \text{AnsIdentify}}
    \label{AnsIdentify}
    \KwIn{Dataset: $\mathcal{D}$; Target Variable: $T$; The Set of Variables to be Searched: $S$; Correlation Coefficient Matrix: $\text{corMat}$; Relevance threshold: $\theta_{R}$; Independence threshold: $\theta_{I}$; Directionality threshold: $\theta_{D}$.}
    \KwOut{Dictionary of Ancestor Sets of Outcome Variables: $An(y)$.}
    
    initialization: search = True; $T = y$; $S = \{X,t\}$; $An(y) = \emptyset$; exSet = $\emptyset$\\
    
    \While{search}{
        parSets = $\emptyset$;\ chiSets = $\emptyset$\\
        \For{$T_{i} \in T$}
        {
        parSet, ChiSet = PCIdentify($\mathcal{D}$,$T_{i}$,$S$,$\text{corMat}$,$\theta_{R}$,$\theta_{I}$,$\theta_{D}$,$\text{exSet}$)\\
        parSets = parSets $\cup$ parSet\\
        chiSets = chiSets $\cup$ ChiSet\\
        $An(y)[T_{i}]$ = parSet\\
        }
        \eIf{parSets == $\emptyset$}{
            search = False
        }{
            exSet = exSet $\cup$ chiSets\\
            $T$ = parSets
        } 
    }
    \Return $An(y)$
\end{algorithm}

Algorithm \ref{PCIdentify} constitutes the core functionality within ASI algorithm, aimed at identifying and distinguishing the parent and child nodes of any given variable. Lines 3-18 employ the GCM\cite{shah2020hardness} model, grounded in the theory of conditional independence\cite{fu2008fast,spirtes2001causation} on causal diagrams, to recognize the parent and child nodes of the target variable. Subsequently, lines 19-27 implement the ANM\cite{hoyer2008nonlinear} model, based on the theory of causal asymmetry, to explicitly differentiate between parent and child nodes. In line with the assertion from \cite{yan2020effective}: "If a conditioning set of more than three variables does not render two variables independent, then it is highly probable that these two variables are directly related", lines 14-16 (size $\geq$ 4) operationalizes this judgment to reduce the complexity of the ASI algorithm.

\begin{algorithm}
    \caption{The algorithm procedure of \text{PCIdentify}}
    \label{PCIdentify}
    \KwIn{Dataset: $\mathcal{D}$; Target Variable: $T$; The Set of Variables to be Searched: $S$; Correlation Coefficient Matrix: $\text{corMat}$; Relevance threshold: $\theta_{R}$; Independence threshold: $\theta_{I}$; Directionality threshold: $\theta_{D}$; The Set of Excluded variables: exSet.}
    \KwOut{The Set of Parents and Children of the Target Variable: parSet, ChiSet.}
    
    initialization: parSet = $\emptyset$; ChiSet = $\emptyset$ \\
    Adj = $\{X^{i}:\text{corMat}_{X^{i}}^{T} \geq \theta_{R} \}$;\ Adj = Adj $\setminus$ $T$ $\setminus$ exSet;\ sortAdj = sorted($\text{corMat}_{\text{Adj}}^{T}$)\\
     \For{$\text{Adj}_{i} \in \text{sortAdj}$} % adj
        {
        CAdj = Adj $\setminus$ $\text{Adj}_{i}$;\ size = 1;\ test = True\\
        \While{test}{
            \For{$\text{CAdj}_{i} \in \text{combinations(CAdj, size)}$}
            {
            PValue = GCM($T$, $\text{Adj}_{i}$, $\text{CAdj}_{i}$)
            
            \If{PValue $\geq$ $\theta_{I}$}{
                Adj = CAdj\\ search = False
            }
            }
            size = size + 1\\
            \If{(size $\geq$ 4) or (size $\geq$ len(CAdj))}{
                search = False
            }
        }
        }

    \For{$\text{PC}_{i} \in \text{Adj}$} % pc
        {
        PFoward, PBack = ANM($T$, $\text{PC}_{i}$)\\
        \If{(PFoward $\geq$ $\theta_{D}$) and (PBack $\leq$ $\theta_{D}$)}{
                ChiSet = ChiSet $\cup$ $\{\text{PC}_{i}\}$
            }
        \If{(PFoward $\leq$ $\theta_{D}$) and (PBack $\geq$ $\theta_{D}$)}{
                parSet = parSet $\cup$ $\{\text{PC}_{i}\}$
            }
        }
    \Return parSet, ChiSet   
\end{algorithm}
\subsection{The Overview of the GCI Framework}
\begin{figure}
\centering
\includegraphics[width=0.98\textwidth]{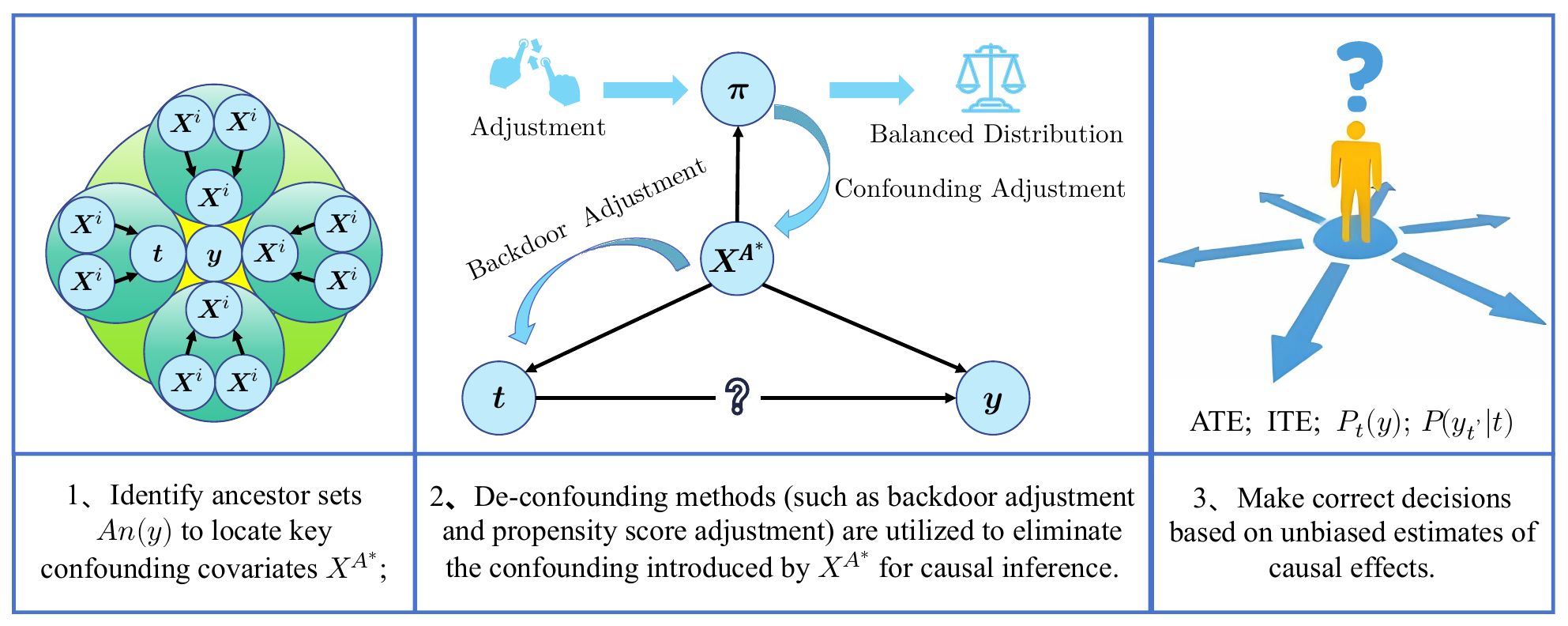}
\caption{The Overview of the GCI Framework.}
\label{Frame}
\end{figure}

Drawing on the conclusions derived in Section 3.2 (the common ancestral root nodes of treatment and outcome variable are key confounding covariates), and integrating the ASI algorithm from Section 3.3 with the model for eliminating confounding influences, this section presents a GCI framework for cross-sectional observational data, as illustrated in Fig. \ref{Frame}.

The overview depicted in Fig. \ref{Frame} delineates the fundamental process of decision-making based on the GCI framework: For observational data, one initially employs ASI algorithm to identify the ancestral set of the outcome variable and to pinpoint the key confounding covariates $X^\textit{A*}$; This is further complemented by de-confounding methods (such as backdoor adjustment, propensity score adjustment, etc.) to facilitate unbiased causal inference, which encompasses both group-level effect estimation ($P_{t}(y)$ or Average Treatment Effect, ATE) and individual-level counterfactual inference ($P(y_{t^{'}}|t)$ or Individual Treatment Effect, ITE); Ultimately, informed decisions are made based on the causal effect inferences obtained, with practical applications that include, but are not limited to, the assessment of regulatory relationships in bioinformatics systems, the evaluation of critical prognostic physiological indicators in clinical research, and the refinement of recommendation strategies in recommendation systems.
\section{Experiments}
\subsection{Datasets}
\begin{figure}[ht]
\centering
\includegraphics[width=0.65\textwidth]{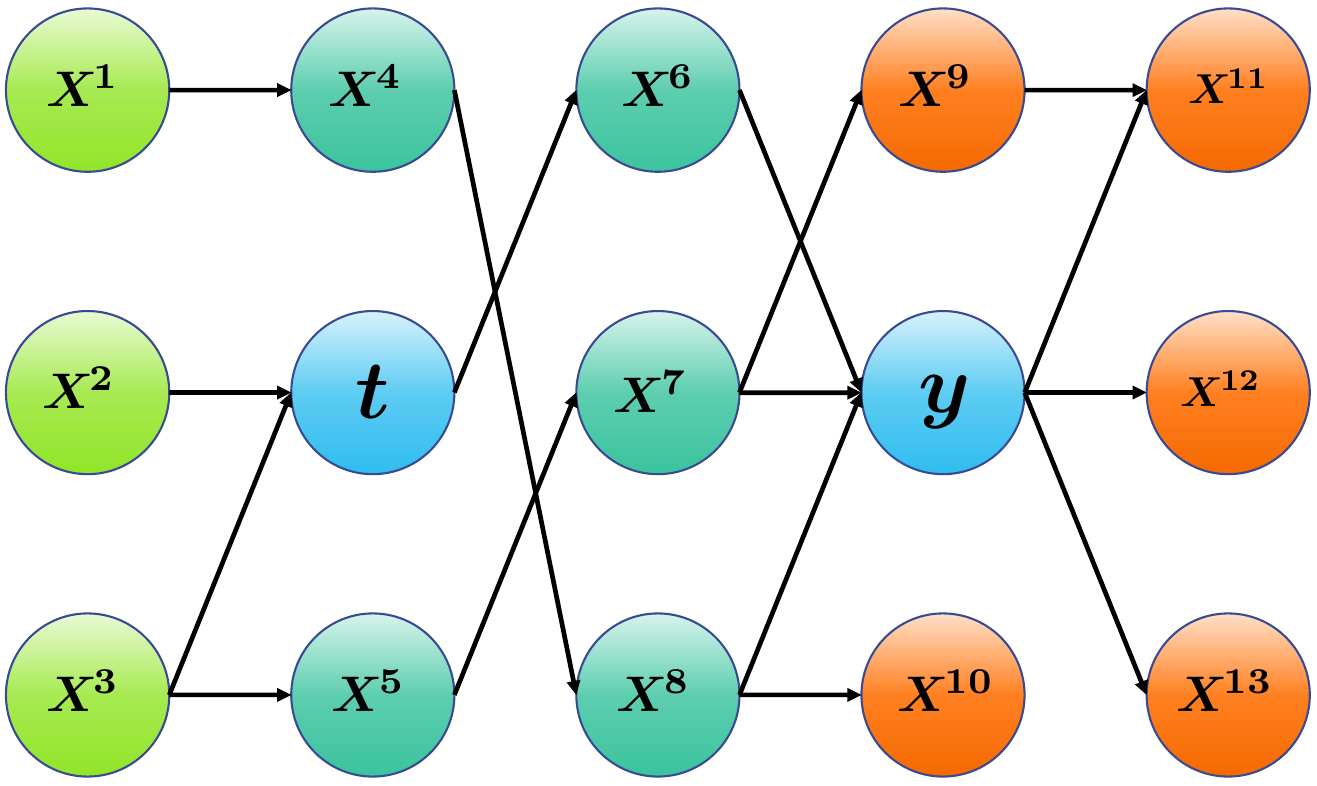}
\caption{The causal diagram corresponding to the synthetic dataset.}
\label{CASAULexp}
\end{figure}

In practice, researchers often encounter a situation where they can only observe the factual outcome of a particular treatment, while the corresponding counterfactual outcomes remain unknown. Additionally, identifying the precise components of the covariates in real-world datasets can be challenging. To address these limitations, previous studies have utilized synthetic or semi-synthetic datasets\cite{Hill2011Bayesian,dehejia1999causal,NIPS2017Latent,johansson2018learning}. Informed by these research endeavors, this paper has generated synthetic data guided by the intricate causal diagram illustrated in Fig. \ref{CASAULexp}.

Based on the causal diagram depicted in Fig. \ref{CASAULexp}, the synthetic dataset $\mathcal{D}^{S}$ consists of three components: $\mathcal{D}^{S}= \{t,X,y\}$, where $X=\{X^{i}\}_{i=1}^{13}$. The variables $X^{1}$, $X^{2}$, and $X^{3}$ are root variables and randomly generated from a Normal distribution, as indicated in Eq.~\eqref{equ17}.
\begin{equation}
\label{equ17}
X^{1}, X^{2}, X^{3} \sim N(0,1)
\end{equation}

Subsequently, the remaining non-root variables, namely $V$, are sequentially generated based on their respective parent variables $Pa(V)$, as depicted in Eq.~\eqref{non-root}. It should be emphasized that all parameters, denoted as $W$, follow a uniform distribution between $[0.5,2]$. Moreover, the nonlinear functions $f$ are specified as the logistic function $Sigmoid$.
\begin{equation}
\label{non-root}
V = \sum_{Pa(V)}f(Pa(V)\cdot W) +  \epsilon_{V}, \epsilon_{V} \sim N(0,1)
\end{equation}

The dataset $\mathcal{D}^{S}$ is comprised of a sample size of 1,000, generated according to the mechanism illustrated in Fig. \ref{CASAULexp}. Given the intricate causal interdependencies between covariates $X$, treatment variables $t$, and outcome variables $y$, directly quantifying the causal effect of $t$ on $y$ from $\mathcal{D}^{S}$ represents a formidable challenge.
\subsection{Metrics} 
To illustrate the accuracy of GCI framework, the error of the marginal treatment effect function (MTEF) is used to measure the causal inference performance in the case of the continuous treatment\cite{Kreif2015Evaluation}.As demonstrated in Eq. (\ref{equ3}), the MTEF indicates the causal effect of a perturbation at a particular treatment level on the expected counterfactual outcome for all samples. In the other word, the MTEF captures the marginal change in the outcome variable caused by the treatment variables at a particular level in a differential form.
\begin{equation}
\label{equ3}
{\rm MTEF}(t) = \frac{\mathbb{E}[y_{i}(t)] - \mathbb{E}[y_{i}(t - \Delta t)]}{\Delta t}
\end{equation}
Next, we measure the accuracy of counterfactual inference by comparing the ${\rm MTEF}^{pre}$ predicted by the comparison model with the true ${\rm MTEF}^{true}$, that is the rooted mean squared error (RMSE) of MTEF shown in Eq.~(\ref{equ21}). 
\begin{equation}
\label{equ21}
\epsilon_{{\rm MTEF}} = \sqrt{\frac{1}{n}\sum_{i=1}^{n}({\rm MTEF}^{true}_{i}-{\rm MTEF}^{pre}_{i})^{2}}
\end{equation}

\subsection{Experimental Details}
In this paper, the proposed model is compared with various state-of-the-art counterfactual inference models for continuous treatment variables that are based on reweighting or deep networks:  including ICPW\cite{2000Marginal} , GBM\cite{zhu2015boosting}, CBGPS and npCBGPS\cite{2018Covariate},  DRNets\cite{2019LearningC}, DRL\cite{zhao2023deconfounding}.

The datasets for the aforementioned four scenarios are divided into training/test sets according to the percentages of 80/20. Further, as obtaining consistent causal conclusions is one of the primary objectives of causal inference, we treat sampling data greater than eighty percent of the quantile of the treatment variables as test set and the rest as training set. This partitioning enables us to measure the generalization performance of the comparison model by evaluating its ability to perform well outside the training domain. 

To ensure a fair comparison of the comparison models, a systematic grid search approach is adopted to select the optimal hyperparameters. This entails selecting the best hyperparameters for each model from a predefined range based on its performance on the validation sets. Subsequently, we evaluate the chosen models 100 times to record the mean and standard error of the evaluation metrics. For the above evaluation metrics, we give the values in contexts including training and test sets.

\subsection{Results}
\begin{figure}[ht]
\centering
\includegraphics[width=0.98\textwidth]{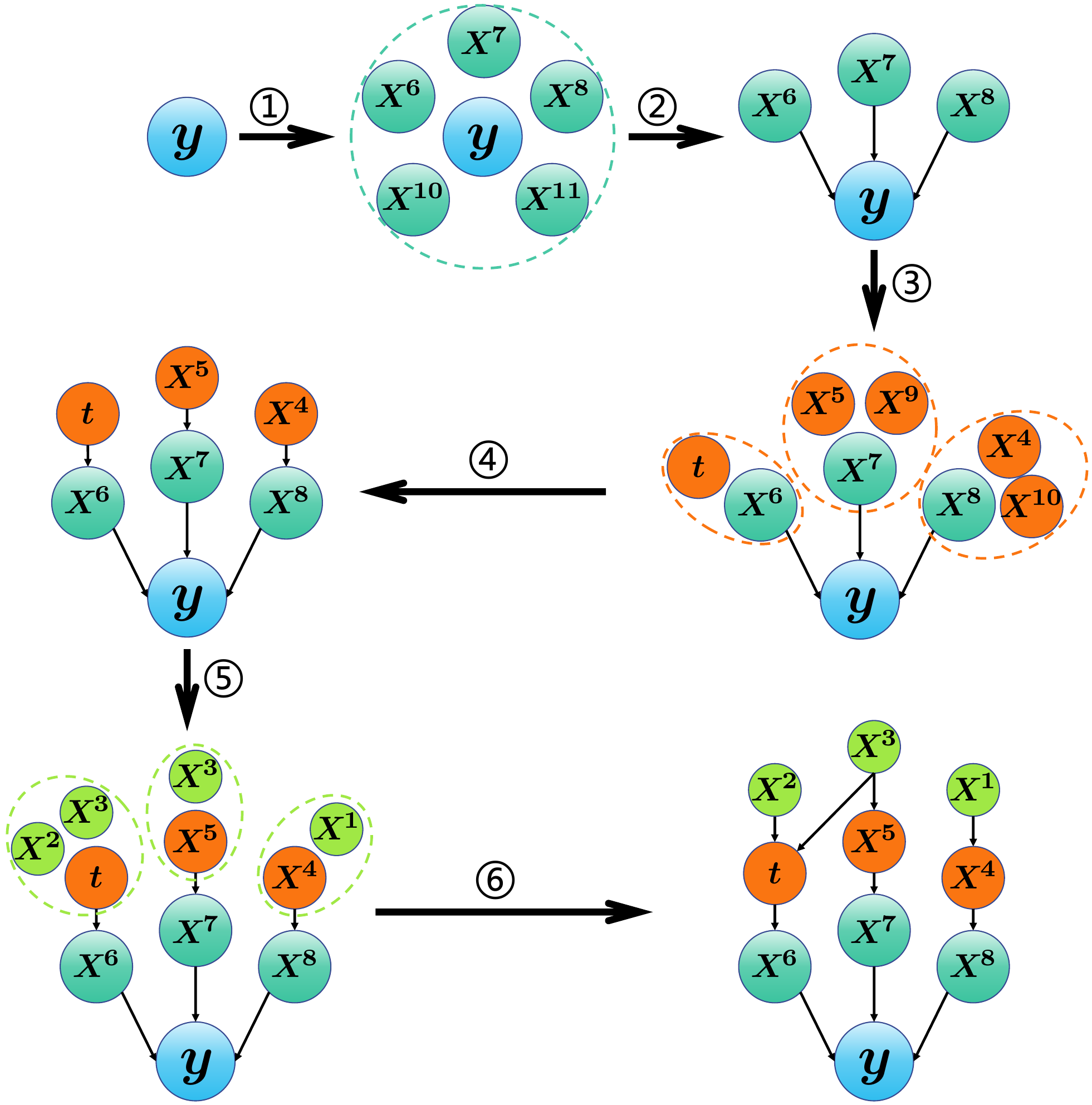}
\caption{The flowchart illustrating the process of identifying the ancestor set of outcome variable by the ASI algorithm.}
\label{Alg2}
\end{figure}

The first step in the GCI framework involves identifying the ancestral set of the outcome variable using the ASI algorithm. Fig. \ref{Alg2} presents the implementation process of this algorithm on dataset $\mathcal{D}^{S}$. It is apparent that the ASI algorithm initiates the iteration from $y$, and by recursively invoking the parent node identification algorithm shown in Algorithm \ref{PCIdentify}, it progressively discerns the complete ancestral set $An(y)$ of $y$. In this process, the range of variables pending identification diminishes incrementally, significantly accelerating the identification speed of the ancestral set. Concurrently, based on the definition of the ancestral set $An(y)$ and common root ancestor set $X^\textit{A*}$ of the treatment and outcome variable, in this case, the set $X^\textit{A*}$ has been identified as $X^\textit{A*} = \{X^{3}\}$.

Furthermore, to demonstrate the improvement of counterfactual inference via $X^\textit{A*}$ learned by GCI framework, we independently train various prevalent models with identical hyperparameters on both the raw data $\{X,t,y\}$ and the augmented data $\{X^\textit{A*},t,y\}$. The inferential results of these models are subsequently reported, as depicted in Table~\ref{MTEF}. It is evident that the GCI framework can significantly enhance the predictive accuracy of various counterfactual inference methodologies. Additionally, the GCI framework demonstrates greater stability in its performance on the test set. Beyond accuracy, another significant advantage of GCI framework over other comparative methods is its enhanced interpretability, which benefits from the initial identification of the ancestral set of the outcome variable.
\begin{table}[ht]
\centering
\caption{The performance of estimating $\epsilon_{{\rm MTEF}}$ with various state-of-the-art counterfactual inference models on synthetic datasets. The term "GCIF" denotes the incorporation of the learned $X^\textit{A*}$ into the corresponding counterfactual inference model.}
%\resizebox{0.6\linewidth}{!}{
\begin{tabular}{|l|cc|cc|}
\hline
\multirow{2}{*}{{\bfseries Methods}} &  \multicolumn{4}{|c|}{{\bfseries Metrics: $\epsilon_{{\rm MTEF}}$}}\\
\cline{2-5}
& Training  & With GCIF &  Test  & With GCIF \\
\hline
ICPW & $1.58\pm0.05$ & \bm{$1.48\pm0.03$} & $2.13\pm0.13$ & \bm{$1.83\pm0.09$}  \\
GBM & $1.57\pm0.05$ & \bm{$1.48\pm0.03$} & $2.07\pm0.08$ & \bm{$1.83\pm0.09$} \\
CBGPS & $1.63\pm0.06$ & \bm{$1.49\pm0.04$} & $2.06\pm0.08$ & \bm{$1.87\pm0.09$} \\
npCBGPS & $1.58\pm0.05$ & \bm{$1.48\pm0.04$} & $2.03\pm0.13$ & \bm{$1.79\pm0.06$} \\
DRNets & $1.64\pm0.08$ & \bm{$1.60\pm0.07$} & $1.73\pm0.08$ & \bm{$1.69\pm0.07$} \\
DRL & $1.58\pm0.07$ & \bm{$1.51\pm0.05$} & $1.79\pm0.11$ & \bm{$1.65\pm0.06$} \\
\hline
\end{tabular}
%}
\label{MTEF}
\end{table}
\section{Conclusion}
The research on causal inference methods for observational data has been highly regarded and plays a crucial role in fields such as economics, healthcare, and recommendation systems. The primary objective of these studies is to eliminate the influence of confounding factors. Most relevant methods default to assuming that all covariates are confounders or make naive assumptions about covariates. However, in practice, covariates are often high-dimensional and exhibit complex causal relationships, making it difficult to identify the key confounding covariates and limiting the practical significance of these methods. This paper proposes the GCI framework specifically designed for cross-sectional observational data, which targets the identification of key confounding variables and mitigates their impact on causal inference to relax the naive assumptions and provide practical feasibility. We first derive, from a theoretical perspective based on the Markov property on causal diagrams, that the key confounding covariates in causal effect estimation are the common root ancestors of the treatment and outcome variables. Based on the conditional independence properties and causal asymmetry between causal variables, we then design an ASI algorithm to acquire the key confounding covariates. Finally, we combine this algorithm with the de-confounding inference methods to construct the GCI framework. Extensive experiments on synthetic datasets demonstrate that the proposed GCI framework can effectively identify the key confounding covariates and significantly improves the precision, stability, and interpretability of causal inference tasks. In the future, The exploration of more efficient ancestral set identification algorithms is a promising direction worthy of further investigation.


\bibliographystyle{cas-model2-names}
% \bibliographystyle{model1-num}

\bibliography{mybib}
\end{document}